\documentclass[a4paper,12pt]{article}

\usepackage[left=25mm, right=25mm, bottom=35mm, top=35mm]{geometry}

\usepackage[T2A]{fontenc}
\usepackage[utf8]{inputenc}
\usepackage[english]{babel}
\usepackage[pdftex]{graphicx}

\usepackage{amsmath,amssymb,ntheorem}
\usepackage{multicol}
\usepackage{colortbl} 

\usepackage{geometry}
\usepackage{epsf}
\usepackage{amsfonts}
\usepackage{amsmath}
\usepackage{amssymb}
\usepackage{latexsym}
\usepackage{units}
\usepackage{textcomp}
\usepackage{wrapfig}
\usepackage{booktabs}
\usepackage{hyperref}

\usepackage{xcolor}
\hypersetup{
    colorlinks,
    linkcolor={red!50!black},
    citecolor={blue!50!black},
    urlcolor={blue!80!black}
}

\everymath{\displaystyle}
\usepackage{color}
\everymath{\displaystyle}

\renewcommand{\phi}{\varphi}

\usepackage{url}

\newcommand{\descrfnt}{\fontfamily{ppl}\selectfont\small}

\begin{document}
~
\vspace{-30mm}
\begin{flushleft}
\footnotesize  
\rmfamily      
\begin{tabular}[c]{l}
\rowcolor[gray]{.9}This is a pre-print of an article published in Eur.\hspace{0.5mm}Phys.\hspace{0.5mm}J.\hspace{0.5mm}C\hspace{0.4mm}(2018)\hspace{0.4mm}78:\hspace{0.2mm}92.\\
\rowcolor[gray]{.9}The final authenticated version is available online at: \href{https://doi.org/10.1140/epjc/s10052-018-5566-x}{https://doi.org/10.1140/epjc/s10052-018-5566-x}
\end{tabular}
\end{flushleft}
\vspace{20mm}

\begin{center}
\section*{ \boldmath Search for heavy neutrino in $K^{+} \to \mu^{+} \nu_{H}$ decay}
\end{center}

\begin{center}
{\large 
\textsc{The OKA collaboration}\\
}
\vspace{3mm}

\begin{minipage}{0.9\linewidth}
\center{
  \textsc
  A.S.~Sadovsky, V.F.~Kurshetsov, A.P.~Filin, S.A.~Akimenko, A.V.~Artamonov, 
  A.M.~Blik, V.V.~Brekhovskikh, V.S.~Burtovoy, S.V.~Donskov, A.V.~Inyakin,
  A.M.~Gorin, G.V.~Khaustov, S.A.~Kholodenko,
  V.N.~Kolosov, A.S.~Konstantinov, V.M.~Leontiev, V.A.~Lishin,  M.V.~Medynsky,
  Yu.V.~Mikhailov, V.F.~Obraztsov, V.A.~Polyakov, A.V.~Popov, V.I.~Romanovsky,
  V.I.~Rykalin,  V.D.~Samoilenko,  V.K.~Semenov,
  O.V.~Stenyakin,  O.G.~Tchikilev, V.A.~Uvarov, O.P.~Yushchenko
 }
 \vspace{-4mm}
 \center{\small 
   \textsc{(NRC "Kurchatov Institute"${}^{}_{}{}^{}$-${}^{}_{}{}^{}$IHEP, Protvino, Russia),} 
 }\\
 \vspace{-3mm}
 \center{
  \rmfamily
  V.A.~Duk\footnote{\scriptsize Also~at~University~of~Birmingham,~Birmingham,~United~Kingdom}, 
  S.N.~Filippov, E.N.~Gushchin, 
  A.A.~Khudyakov, V.I.~Kravtsov, 
  Yu.G.~Kudenko\footnote{\scriptsize Also at Moscow Institute of Physics and Technology, Moscow Region, Russia}\footnote{\scriptsize Also at NRNU Moscow Engineering Physics Institute (MEPhI), Moscow, Russia}, 
  A.Yu.~Polyarush
 }\vspace{-4mm}
 \center{\small 
    \textsc{(INR RAS, Moscow, Russia),}
 }\\
 \vspace{-3mm}
 \center{
  \rmfamily
  V.N.~Bychkov, G.D.~Kekelidze, V.M.~Lysan, B.Zh.~Zalikhanov
 }\vspace{-4mm}
 \center{\small 
   \itshape 
   \textsc{(JINR, Dubna, Russia)}\\
 }
\end{minipage}
\end{center}


\vspace{0mm}
\begin{center}
\begin{minipage}{0.09\linewidth}
~
\end{minipage} 
\begin{minipage}{0.70\linewidth}
{ 
  \rmfamily
{\bf Abstract.}
A high statistics data sample of the $K^{+}\to\mu^{+}\nu_{\mu}$ decay was accumulated by the OKA experiment in 2012. 
The missing mass analysis was performed to search for the decay channel $K^{+}\to\mu^{+}\nu_{H}$ 
with a hypothetic stable heavy neutrino in the final state.
The obtained missing mass spectrum does not show peaks which could be attributed to existence of
stable heavy neutrinos in the mass range (220 $< m_{\nu_{H}} <$ 375) MeV$/c^{2}$. 
As a result, we obtain upper limits on the branching ratio and on the value of 
the mixing element $|U_{\mu H}|^{2}$.
}
\end{minipage}
\begin{minipage}{0.09\linewidth}
~
\end{minipage}
\end{center}
\vspace{0mm}

\selectlanguage{english}


%
%
%

\section{Introduction}\label{SectInitro}
After the discovery of the Higgs boson there are no further guideline predictions from 
the Standard Model (SM) remained, hence, searches for a physics beyond the SM in a 
broad range of topics become an actual question. One of the promising research 
directions is inspired by observed neutrino oscillations \cite{SuperKAMIOKANDE, SNO, KamLAND, K2K}
which require non zero neutrino masses 
which, in turn, open possibility for existence of a set of heavy sterile neutrinos in one of the 
SM extensions -- the Neutrino Minimal Standard Model ($\nu$MSM) \cite{nuMSM1,nuMSM2,nuMSM3}. 
Depending on the lifetime of those heavy neutrinos ($\nu_{H}$) experimental approaches can be divided 
into searches for possible decay products of $\nu_{H}$, as, for example, reported in \cite{CERN-PS191} or 
more recently in \cite{vDukISTRA_HNu} and into searches for $\nu_{H}$ with a long lifetime with 
the missing mass approach \cite{KEK-E-089} and, recently, in two high statistics experiments \cite{ArtamonovEtAl-E949}, 
\cite{PublNA62}, 
which reported upper limits on branching for $\nu_{H}$ in a hundred MeV/c$^2$ range. 
\begin{figure}[!h]
\begin{minipage}{0.29\linewidth}
\includegraphics[width=0.89\linewidth]{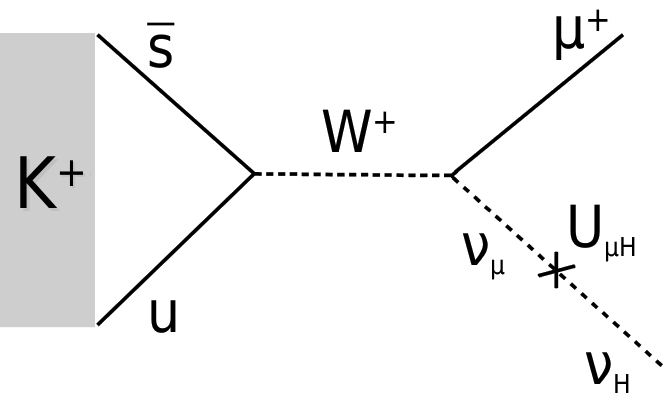}
\end{minipage} 
\begin{minipage}{0.69\linewidth}
\caption[Production of heavy sterile neutrino in a $K^{+}$ decay.]{
  \descrfnt   
  Production of a heavy sterile neutrino in the $K^{+}$ decay.
  \label{FigHeavySterileNeutrinoKmuHnu}
}
\end{minipage}
\end{figure}  
To contribute to the latter approach we analyzed a large data sample 
of $K^{+} \to \mu^{+} \nu$ recorded in 2012 by the OKA collaboration at IHEP-Protvino to search for a 
possible process $K^{+} \to \mu^{+} \nu_{H}$, see Fig.~\ref{FigHeavySterileNeutrinoKmuHnu}, 
where $\nu_{H}$ stands for one of the expected sterile 
$\nu_{H}$ according to the $\nu$MSM and $U_{\mu H}$ represents mixing element 
between SM muon neutrino and $\nu_{H}$. 

\section{Separated kaon beam and OKA experiment}\label{SectExpOverview}
The OKA\footnote{From abbreviation for {\it{``experiments {\bfseries O}n {\bfseries KA}ons''}}.} experiment 
makes use of a secondary hadron beam of the U-70 Proton Synchrotron of 
NRC "Kurchatov Institute"-$^{}$IHEP, Protvino,
with enhanced fraction of kaons obtained by RF-separation 
with Panofsky scheme \cite{OkaSecondaryBeam}. 
Corresponding deflectors are two superconducting Karlsruhe-CERN cavities used at SPS \cite{CERN_KaonSeparator} 
and which were donated by CERN to IHEP in 1998. The cavities are cooled by superfluid He provided by 
the dedicated IHEP-build cryogenic system \cite{CryiogenicsForOKA}. 
The design is optimized for the momentum of 12.5 and 17.7 GeV/c, the achieved fraction of kaons is up 
to 20\% depending on the beam momentum, with the intensity of about $5\times 10^{5}$ kaons per U-70 
spill of 3 seconds duration. The r.m.s.~width of the momentum distribution is
estimated to be 1.5\%.
%
%

\begin{figure}[!ht]
\includegraphics[width=1.0\textwidth]{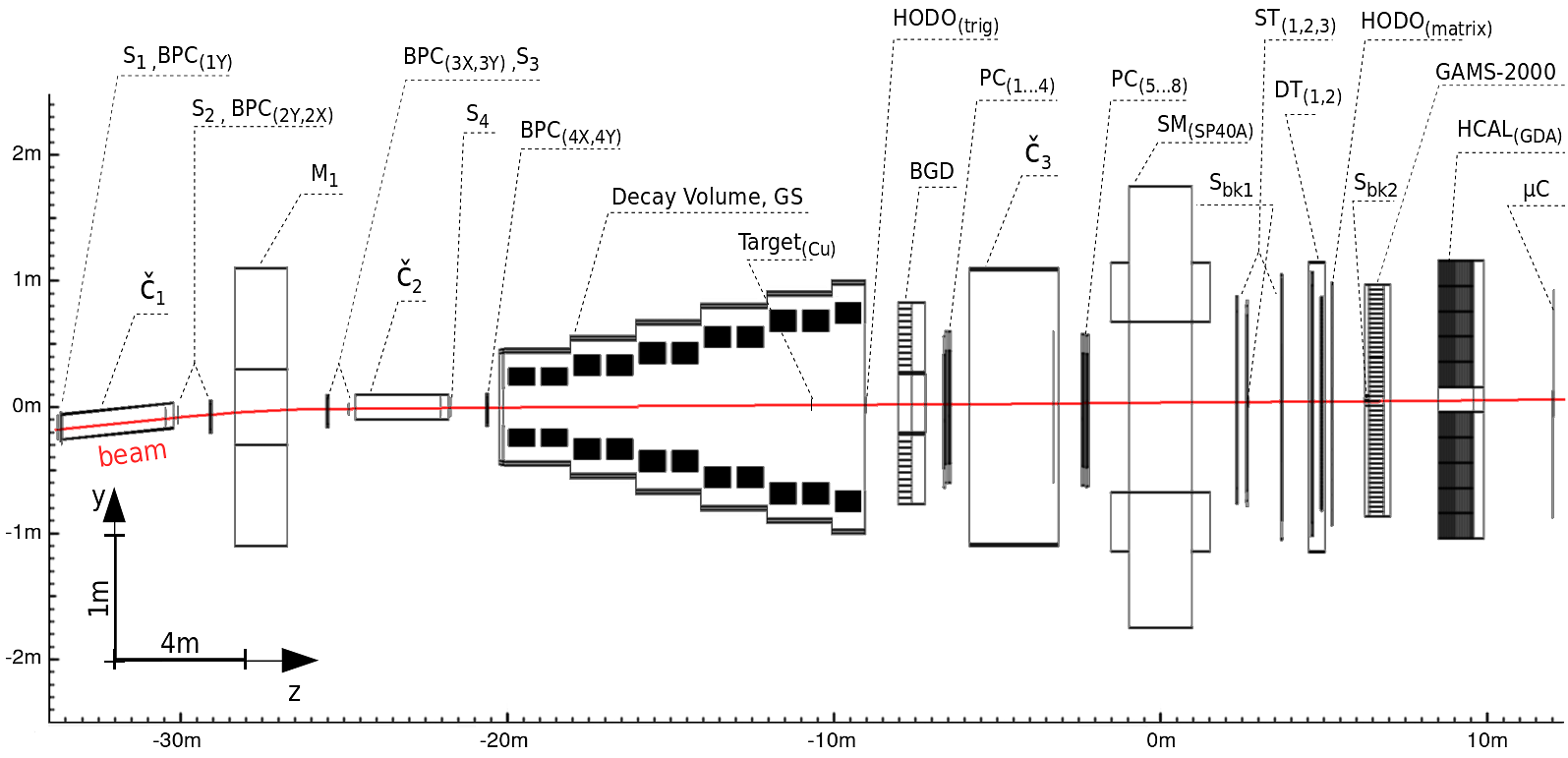} 
\caption[OKA setup.]{
    \descrfnt
    Schematic elevation view of the OKA setup, see text for details.
}
\label{FigOkaSetup}
\end{figure}

The OKA setup, Fig.~\ref{FigOkaSetup}, is a magnetic spectrometer complemented by electromagnetic
and hadron calorimeters and a Decay Volume. First magnet {\small M$_{1}$} with surrounding 1~mm pitch PC's 
({\small BPC$_{(1Y)}$, BPC$_{(2Y,2X)}$}, {\small BPC$_{(3X,3Y)}$}, {\small BPC$_{(4X,4Y)}$} of $\sim$1500 channels 
in total \cite{BPCref}) 
serves as a beam spectrometer.
It is supplemented by two threshold Cherenkov counters {\small \v{C}$_{1}$, \v{C}$_{2}$} for kaon selection 
and by beam trigger scintillation counters 
{\small S$_{(1)}$}, {\small  S$_{(2)}$}, {\small S$_{(4)}$}, each of
200$\times$200$\times$1~mm$^{3}$, and a thicker one, 60$\times$85$\times$6~mm$^{3}$, delivering timing, {\small S$_{(3)}$}.
The 11~m long {\small Decay Volume (DV)} filled with helium contains 11 rings of guard system ({\small GS}),
which consists of 670 Lead-Scintillator sandwiches (20 layers of 1.5mm:5mm each) 
with WLS readout grouped in 300 ADC channels. To supplement {\small GS}, 
a gamma detector ({\small BGD}, made of $\sim$1050 5$\times$5$\times$42~cm$^3$ lead glass blocks \cite{BGDref1982}), 
located behind the {\small DV} is used as a veto at large angles, 
while low angle particles pass through a central opening. 
The wide aperture 200$\times$140 cm$^2$ spectrometric magnet, 
{\small SM$_{(SP40A)}$}, with a field integral of $\sim$1~Tm serves as a spectrometer for the charged 
decay products together with corresponding tracking chambers: 5k channels of 2~mm pitch PC's ({\small PC$_{1,...,8}$}), 
1k channels of 9~mm diameter straw tubes {\small ST$_{(1,2,3)}$} and 300 channels of 40~mm diameter drift tubes {\small DT$_{1,2}$}.
The matrix hodoscope {\small HODO$_{(matrix)}$} is composed of 252 12$\times$12$\times$1.5~cm$^3$
scintillator tiles with WLS+SiPM readout. It is used to improve time resolution and to link $x$--$y$ projections of a track.
Two scintillator counters {\small S$_{bk1}$, S$_{bk2}$} (80~mm and 90~mm in diameter with a thickness of 3.9~mm and 5~mm) 
serve to suppress undecayed beam particles.
At the end of the OKA setup there are two calorimeters: electromagnetic ({\small GAMS-2000} of $\sim$ 2300 3.8$\times$3.8$\times$45~cm$^3$ 
lead glass blocks \cite{GAMSref1985}) and a hadron one ({\small HCAL$_{(GDA)}$} of 100 20$\times$20$\times$108~cm$^3$
iron-scintillator sandwiches with WLS plates readout \cite{GDAref1986}) and, finally, four partially overlapping 
muon counters {\small {$\mu$}C} (1$\times1$~m$^2$ scintillators with WLS fibres readout) behind the 
{\small HCAL}. 

\section{Search for heavy neutrinos}
A search for $K^{+}\to\mu^{+}\nu_{H}$ decay is done with the data set accumulated in November 2012 
run\footnote{Half of this run was dedicated to $K^{+}$ -Cu scattering experiment, 
hence during part of data taking a copper target was installed at the end of {\small DV}.} 
with a 17.7 GeV/c beam momentum. Two prescaled triggers are used.
The first one selects beam kaons which decay inside the OKA setup, 
${\small \tt Tr_{Kdecay} = S_{1} \cdot S_{2} \cdot S_{3} \cdot S_{4} \cdot 
\check{C}_{1} \cdot \overline{\check{C}}_{2} \cdot \overline{S}_{bk} }$, prescale factor is 1/10,
while the second one, ${\small \tt Tr_{K\to\mu X} = Tr_{Kdecay} \cdot {\mu}C}$, includes
additionally muon counters {\small $\mu$C} and is prescaled by 1/4.
The beam intensity (S$_{1}\cdot$S$_{2}\cdot$S$_{3}\cdot$S$_{4}$) was $\sim 2\cdot 10^{6}$ per spill, 
the fraction of kaons in the beam is $\sim$ 12.5\%, i.e.~the kaon intensity is  $\sim$250k/spill. 
The total number of $\sim 1.6\times 10^{10}$ events with
kaon decays are logged. 

\subsection{Event selection}
To select $K^{+} \to \mu^{+}$~$\nu$ decay channel in off-line analysis a set of requirements is applied:\\ 
-- events with single beam track and single secondary track are selected;\\
-- a single secondary track segment after the {\small SM} magnet is present in the event
and it is well matched to showers of the muon type (i.e.~one or two adjacent cells
with the MIP energy deposition) in both {\small GAMS-2000} and {\small HCAL} calorimeters;\\
-- sufficient number of points on all the track segments is present to optimize the
missing mass resolution;\\
-- the momentum of kaon is consistent with that delivered by beam settings of 
$\approx 17.7$~GeV/c, while the required momentum of a secondary muon is below $16.4$ GeV/c;\\
-- to ensure good decay vertex reconstruction and also to suppress events in which kaon 
decays behind the {\small DV}, there is a requirement of 3~mrad minimal angle
between the beam and the secondary track, and a requirement for the minimal
distance between the beam and the secondary track to be $< 1$~cm;\\
-- the decay vertex is inside {\small DV}, and is further restricted to be 2$\sigma$ (of $z$-vertex resolution) 
from  the {\small DV} entrance and the position of Cu-target;\\
-- other decay channels are suppressed by requiring the total 
energy deposition in {\small GS} and {\small BGD} to be below 50~MeV/c$^2$ and 100~MeV/c$^2$, respectively;\\
-- the total energy deposition in GAMS-2000 and HCAL should be consistent with that of a single muon.

After applying these cuts, $26\times10^{6}$ $K^{+}\to\mu^{+}\nu$ decays were selected for subsequent analysis.

\subsection{Signal and background studies}\label{SectSigBg}
\begin{figure}[!ht]
\includegraphics[width=1.0\textwidth]{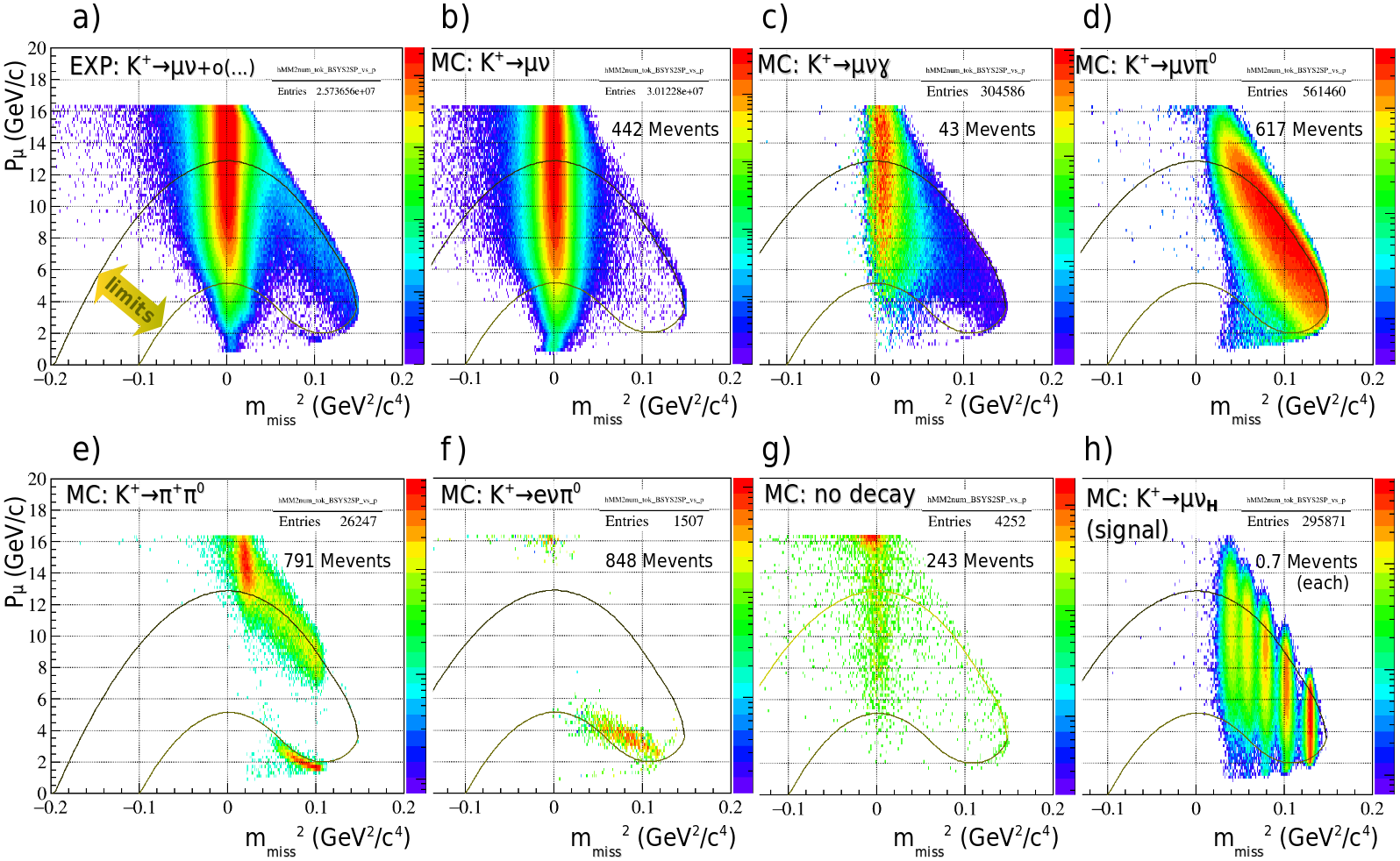} 
\caption[Distributions of $p_{\mu}$~vs.~$m^2_{miss}$ for main background sources.]{
    {\descrfnt
    The distribution of $p_{\mu}$~vs.~$m^2_{miss}$ obtained from the analysis of the experimental data (plot $a$) 
    in comparison with Monte-Carlo simulation for the background channels (plots $b$, ..., $g$) 
    and for the set of signals (plot $h$) for $m_{\nu_{H}} = \{200, 240, 280, 320, 360\}$~MeV/c$^2$.
    Note logarithmic scale for the third dimension. Only events which passed the selection cuts
    are shown. In case of MC, the initial statistics is indicated.
    }
}
\label{SignalAndBg_ps_vs_mm2}
\end{figure}

A signal from heavy neutrino in $K^{+} \to \mu^{+} \nu_{H}$ may show up itself as a peak in the missing mass distribution, 
$m^2_{\nu} = m^2_{miss} = (p_{K} - p_{\mu})_{i} \cdot (p_{K} - p_{\mu})^{i}$,  $i = 1, 2, 3, 4$. 
We assume that heavy sterile neutrinos are stable\footnote{With respect to the size of 
the OKA setup. Consistency check is provided later in subsection \ref{SigSearchMCbgSubtr}.}. 
The investigation of a possible signal decays and background contributions from kaon decays and from 
kaon scattering and interactions inside the OKA setup 
is done with detailed GEANT-3 simulation with the subsequent off-line reconstruction and analysis. 
Different decay channels simulated using Monte-Carlo are weighted according to corresponding matrix 
elements and branchings \cite{PDG}. The experimental data and main backgrounds, which survived the selections 
cuts are shown in the $(m^2_{miss};p_{\mu})$ plots of Fig.~\ref{SignalAndBg_ps_vs_mm2}.

In the region of low $m^{2}_{miss}$, both $K^{+} \to \mu^{+} \nu_{\mu}$ and
$K^{+} \to \mu^{+} \nu_{\mu} \gamma$ dominate.
In the region of $m^{2}_{miss} \gtrsim$ 0.05 GeV$^2$/c$^4$, the dominant contribution is given 
by $K^{+}\to\pi^{0}\mu^{+}\nu_{\mu}$ decay channel, which can not be excluded by kinematic cuts without 
significant loss in acceptance due to its rather flat distribution in the region of interest 
(see Fig.~\ref{SignalAndBg_ps_vs_mm2}-d).

The $K^{+}\to\pi^{+}\pi^{0}$ decay channel is suppressed by four orders of magnitude, but it is 
responsible for a small peak at $m^{2}_{miss}$ around 0.1~GeV$^2$/c$^4$, which 
should be taken into account. Therefore we limit our acceptance (see Fig.~\ref{SignalAndBg_ps_vs_mm2}-a) 
from the low muon momentum side by a smooth curve excluding the high event density spot 
from $K^{+}\to\pi^{+}\pi^{0}$ at low $p_{\mu}\approx 2$ GeV/c (see Fig.~\ref{SignalAndBg_ps_vs_mm2}-e), 
while at high values of the muon momentum ($p_{\mu}$) we additionally introduce a smooth upper limit 
on $p_{\mu}$ to suppress a tail from badly reconstructed $K^{+} \to \mu^{+} \nu_{\mu}$ events.
The contributions due to a misidentified electron from the $K^{+} \to e^{+} \nu_{e} \pi^{0}$ decay channel 
(Fig.~\ref{SignalAndBg_ps_vs_mm2}-f) and from processes when kaon either 
scatters or interacts while passing the setup (Fig.~\ref{SignalAndBg_ps_vs_mm2}-g) play a minor role. 
\begin{figure}[!h]
\begin{minipage}{0.48\linewidth}
\includegraphics[width=0.98\linewidth]{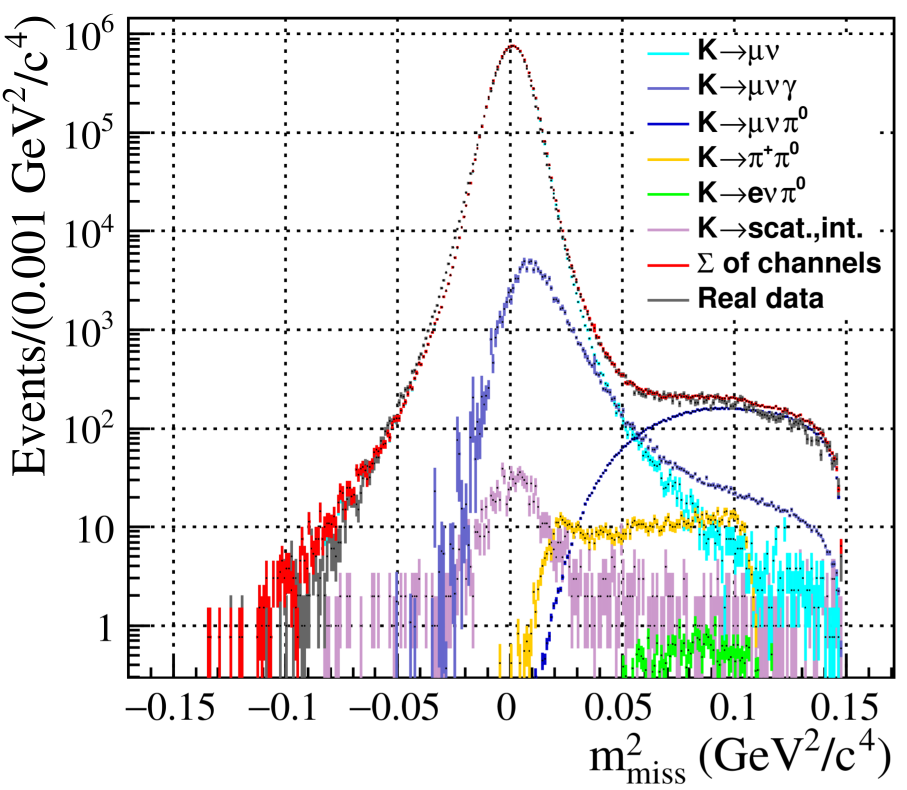} 
\end{minipage}
\begin{minipage}{0.04\linewidth}
~\\
\end{minipage}
\begin{minipage}{0.48\linewidth}
\caption[Missing mass distribution $m^{2}_{miss}$ for experimental data and for different decay channels obtained from simulation.]{
  {
  \descrfnt   
  Missing mass distribution $m^{2}_{miss}$ for the data and MC events inside kinematic region, indicated in
  Fig.~\ref{SignalAndBg_ps_vs_mm2}-a.
  Distribution for the experimental data is shown in gray, 
  dominant decay channels, obtained from MC, are marked by different colors,
  while their sum is depicted in red. The normalization is relative to the
  experimental data.\\
  ~\\
  }
  \label{normalizedToData_SigAndBg_vs_mm2}
}
\end{minipage}
\end{figure}

\begin{figure}[!h]
\begin{minipage}{0.48\linewidth}
\includegraphics[width=0.99\linewidth]{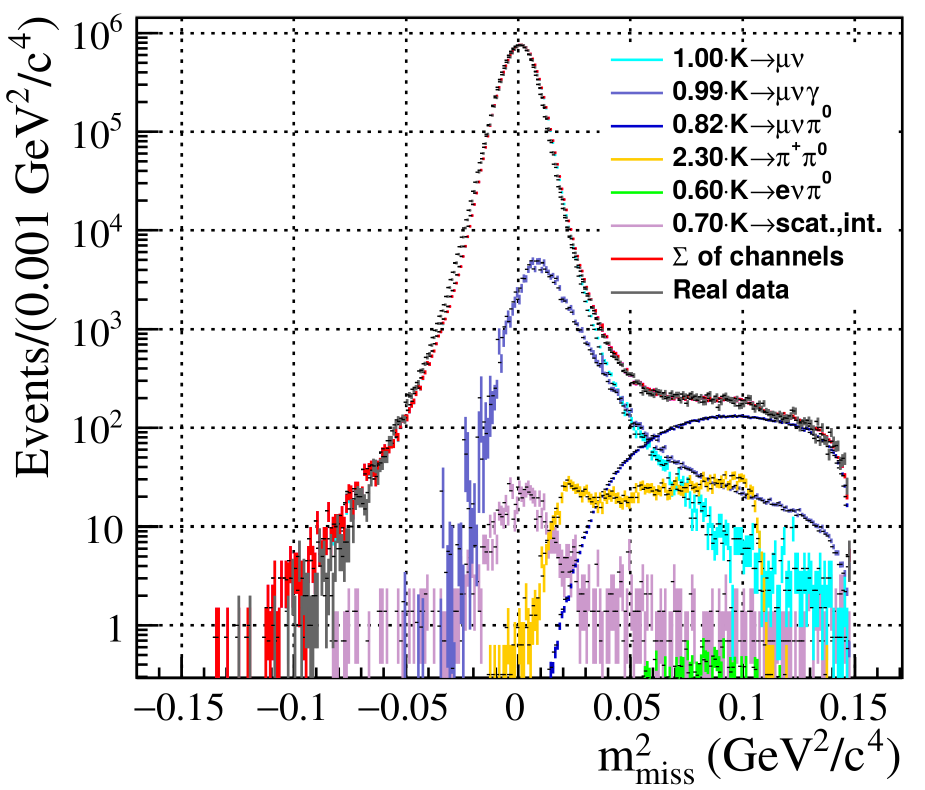} 
\end{minipage}
\begin{minipage}{0.04\linewidth}
~\\
\end{minipage}
\begin{minipage}{0.48\linewidth}
\includegraphics[width=0.99\linewidth]{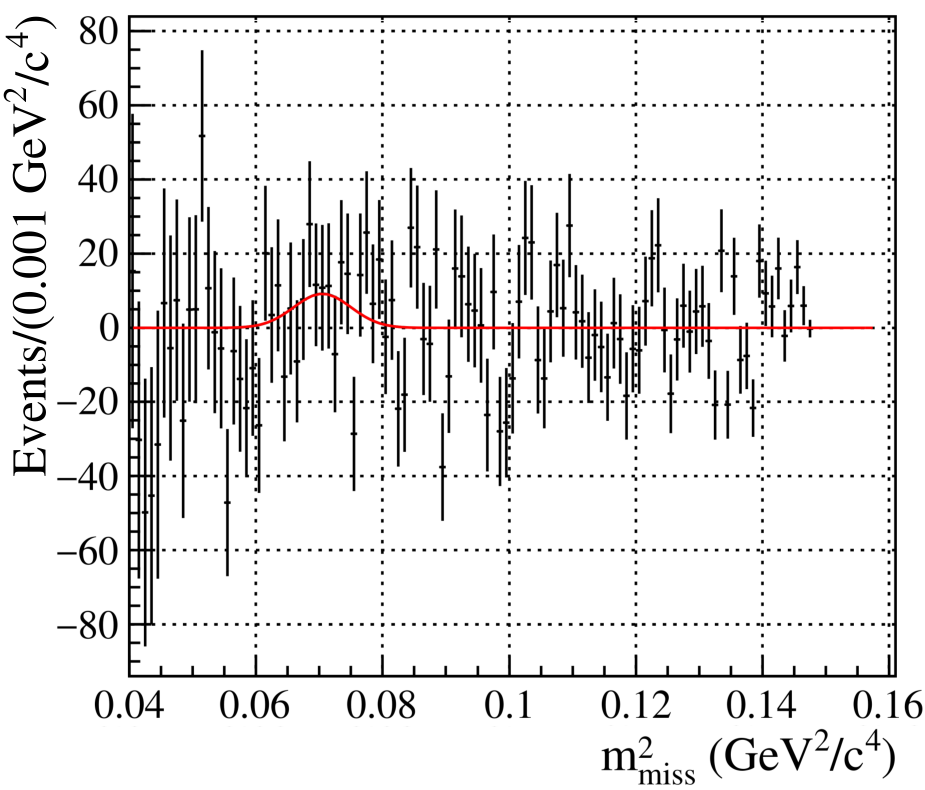} 
\end{minipage}\\
\begin{minipage}{0.47\linewidth}
\caption[Magnitudes of individual MC-simulation terms fitted.]{
  \descrfnt
  Same as Fig.~\ref{normalizedToData_SigAndBg_vs_mm2}, but magnitudes of individual MC-simulation terms 
  are tuned by the fit procedure for the best agreement with the data in the region of $m^2_{miss}>0.05$~GeV$^2$/c$^4$.
  Multiplicative efficiency correction is indicated for each background term.
  \label{normalizedToData_SigAndBgFitted_vs_mm2}
}
\end{minipage}
\begin{minipage}{0.04\linewidth}
~\\
\end{minipage}
\begin{minipage}{0.47\linewidth}
\caption[Residual signal.]{
  \descrfnt
  The residual signal distribution after subtraction of the fitted background 
  (see Fig.~\ref{normalizedToData_SigAndBgFitted_vs_mm2}) 
  from the experimental data in the range of $m^2_{miss}>0.05$~GeV$^2$/c$^4$. 
  An example of the fit for the signal with the certain $m_{\nu_{H}}$
  and appropriate width is shown by the red curve.
  \label{normalizedToData_ResidualSignalAfterFitedBgSubtracted_vs_mm2}
  ~\\
}
\end{minipage}
\end{figure}

Missing mass distribution $m^2_{miss}$ for the data and MC events inside kinematic region,
indicated in Fig.~\ref{SignalAndBg_ps_vs_mm2}-a, is shown in Fig.~\ref{normalizedToData_SigAndBg_vs_mm2}. 
The distribution for the experimental data is shown in gray,
dominant decay channels, obtained from simulation, are marked by different colors with
the explanation in the top-right corner, while their sum is depicted in red. The
normalization is done to the experimental data at $m^2_{miss}=0$. The relative normalization of different
channels is done in accordance with their branching ratios. 
As seen from Fig.~\ref{normalizedToData_SigAndBg_vs_mm2}, there is a reasonable agreement between the experimental and MC data
with some discrepancy in the $m^2_{miss}>0.05$ GeV$^2$/c$^4$ region. 
Since contributions from different background sources 
are strongly suppressed (by orders of magnitude) by our selection criteria, 
one may expect that efficiencies obtained from MC simulations are reproduced 
not at the same level of accuracy as it is the case for the main decay channel, $K^{+}\to\mu\nu_{\mu}$. 
To account for that we do a fit, introducing an additional multiplication constant for each of the 
suppressed background sources. The fit is done within a region of $m^2_{miss}>0.05$ GeV$^2$/c$^4$, 
see Fig.~\ref{normalizedToData_SigAndBgFitted_vs_mm2}. 

\subsection{Signal search with a subtraction of the MC-simulated background}\label{ss_SigSearchMCsubtr}
\begin{figure}[!h]
\begin{minipage}{0.38\linewidth}
\includegraphics[width=0.99\linewidth]{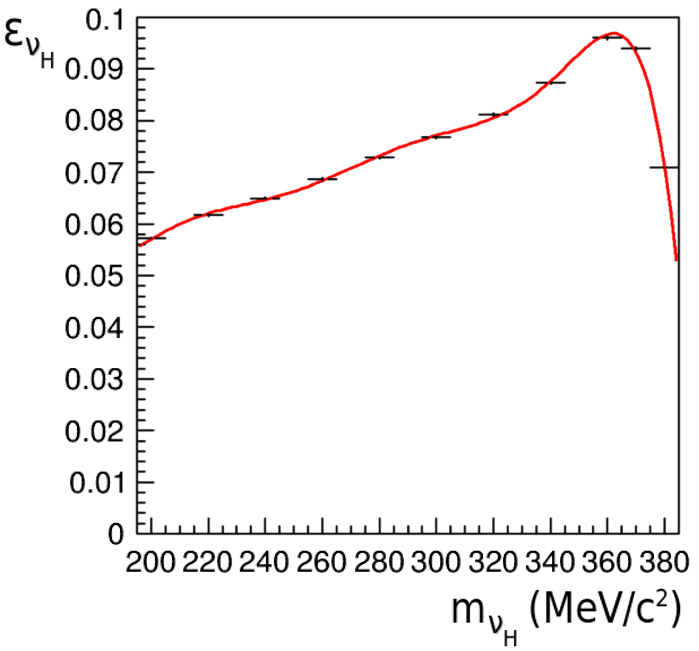} 
\end{minipage}
\begin{minipage}{0.06\linewidth}
~\\
\end{minipage}
\begin{minipage}{0.56\linewidth}
\caption[Signal efficiency dependence on $m_{\nu_{H}}$.]{
  \descrfnt   
  The $m_{\nu_{H}}$ dependence of the total efficiency of the 
  signal obtained for the set of values of heavy neutrino mass.
  The interpolation curve is also shown. 
  ~\\
  \label{efficiencyNuH}
}
\end{minipage}
\end{figure}

As a next step, the obtained background curve is subtracted 
(see Fig.~\ref{normalizedToData_ResidualSignalAfterFitedBgSubtracted_vs_mm2}) for the
subsequent search for heavy neutrinos. For the signal search, a parametrization of the
signal shape is essential. For that, a set of reconstructed heavy neutrino signals was
produced by the MC simulation with a neutrino mass step of 20 MeV/c$^2$.
The signal, as a function of $m^{2}_{miss}$, can be approximated by the Gaussian shape,
with an integral error of $\sim$2\%.
Interpolation for the signal width between the generated set of masses is done with polynomial curves.
The same method is used to produce a total efficiency curve $\varepsilon_{\nu_{H}}$, 
see Fig.~\ref{efficiencyNuH}. 

\begin{figure}[!h]
\begin{center}
\begin{minipage}{0.47\linewidth}
\begin{center}
\includegraphics[width=0.99\linewidth]{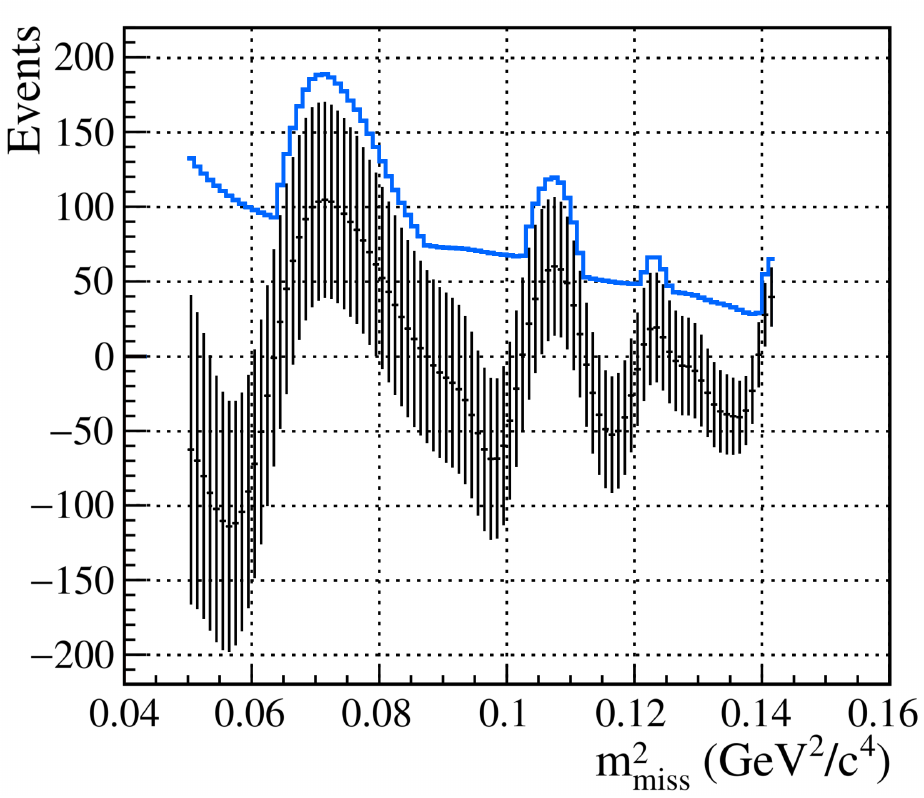} 
\end{center}
\end{minipage}
\begin{minipage}{0.04\linewidth}
~\\
\end{minipage}
\begin{minipage}{0.47\linewidth}
\begin{center}
\includegraphics[width=0.99\linewidth]{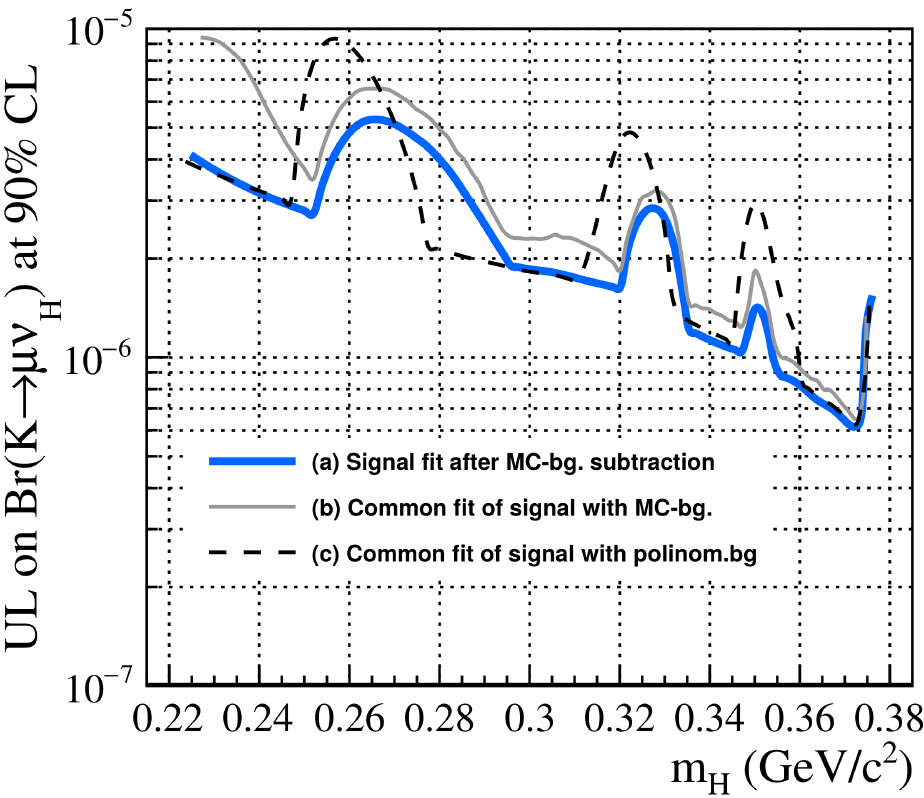}
\end{center}
\end{minipage}\\
\begin{minipage}{0.47\linewidth}
\caption[Magnitudes of individual MC-simulation terms fitted.]{
  \descrfnt
  Number of signal-like heavy neutrino events $N_{\nu_{H}}$ obtained from the fit of the residual 
  distribution for different values of $m^{2}_{\nu_{H}}$ (black points), vertical error bar 
  stands for one standard deviation, $\sigma$. Solid curve corresponds to the upper limit 
  for the number of signal events at 90\% CL.
  \label{EstimNuhNumberBgSubtr_vs_mm2}
}
\end{minipage}
\begin{minipage}{0.04\linewidth}
~\\
\end{minipage}
\begin{minipage}{0.47\linewidth}
\caption[Residual signal.]{
  \descrfnt
  Upper limit for $Br(K^{+}\to\mu\nu_{H})$ at 90\% CL 
  as a function of a heavy neutrino mass for three procedures
  of fit: the bold curve $(a)$ represents procedure described in 
  subsection \ref{ss_SigSearchMCsubtr}, while curves $(b)$ and $(c)$
  are obtained in subsection \ref{ss_SigSearchCommonFitBgWithSig}
  within simultaneous signal and background fit procedures.
  \label{UppLimOnBr90CL_NuhNumberBgSubtr_vs_mm2}
}
\end{minipage}
\end{center}  
\end{figure}

After that, we do series of fits of the residual distribution of 
Fig.~\ref{normalizedToData_ResidualSignalAfterFitedBgSubtracted_vs_mm2} with a Gaussian shape
signal at given mass with an appropriate width i.e.~the only parameter of the fit is
the signal integral which is not restricted by positive values. The obtained number of events 
$N_{\nu_{H}}$ is shown with corresponding errors for each fit in Fig.~\ref{EstimNuhNumberBgSubtr_vs_mm2}.

No indications of signal from $\nu_{H}$ is found in the mass region of
interest. The signal significance does not exceed two standard deviations
in the studied mass interval.
Since the number of events in each bin in the considered $m^2_{miss}$-region is sufficiently high $\sim10^2$
one can apply here the Wald approximation for the likelihood ratio method of upper limits calculation 
for single parameter of interest \cite{gCowan, Wald} which leads to the following single-sided 
upper limit for the signal at 90\%~CL:
{
\begin{center} 
$N^{(Upp.Lim.)}_{\nu_{H}} = 1.28\cdot\sigma + N^{(fit)}_{\nu_{H}}$, ~~~ for $N^{(fit)}_{\nu_{H}}\ge0$,\\
$N^{(Upp.Lim.)}_{\nu_{H}} = 1.28\cdot\sigma$, ~~~~~~~~~~~~~~ for $N^{(fit)}_{\nu_{H}}<0$,\\
\end{center}
}
where $\sigma$ is the standard deviation estimated from the fit, while multiplier of 1.28 
corresponds to one-sided estimate with 90\% CL for the Gaussian case.
The results are shown in Fig.~\ref{EstimNuhNumberBgSubtr_vs_mm2}. 

From that, an upper limit on branching, see bold curve $(a)$ at Fig.~\ref{UppLimOnBr90CL_NuhNumberBgSubtr_vs_mm2}, 
is obtained by normalization to the decay $K^{+} \to \mu \nu_{\mu}$:
$$
Br(K^{+} \to \mu \nu_{H}) = Br(K^{+} \to \mu \nu_{\mu}) \cdot \frac{N_{\nu_{H}}}{\varepsilon_{\nu_{H}}} \cdot \frac{\varepsilon_{\nu_{\mu}}}{N_{\nu_{\mu}}},
$$
where $\varepsilon_{\nu_{H}}$ is obtained from the interpolation function of $m_{miss}$, 
see Fig.~\ref{efficiencyNuH}, $N_{\nu_{\mu}}=12.6\times10^{6}$ is a number of reconstructed 
$K^{+} \to \mu \nu_{\mu}$ events from the experimental data and its efficiency $\varepsilon_{\nu_{\mu}}=0.039$ 
is known from the simulation;
efficiencies and absolute values mentioned here refer to the limited kinematic area 
(defined in Fig.~\ref{SignalAndBg_ps_vs_mm2}-a).

It should be noted that the exclusion of background contributions from
$K^{+} \to e^{+} \nu_{\mu} \pi^{0}$ (Fig.~\ref{SignalAndBg_ps_vs_mm2}-f) and 
from undecayed kaon (Fig.~\ref{SignalAndBg_ps_vs_mm2}-g) 
does not change the obtained results.

\subsection{Upper limits with a common fit of signal and background}\label{ss_SigSearchCommonFitBgWithSig}
As an alternative we do a series of fits of the initial $m^{2}_{miss}$ distribution of Fig.~\ref{normalizedToData_SigAndBgFitted_vs_mm2}
for $m^{2}_{miss}>0.05$~GeV$^2/c^4$ with a sum of Gaussian signal for a given mass and all the MC background sources, again, with additional
multiplication coefficients which now may vary from point to point. Such a procedure is more favorable for the signal amplitude as
compared to that of subsection \ref{ss_SigSearchMCsubtr}, where the background is estimated once before the subtraction and, hence, gives 
more conservative estimate for the upper limit. 
The remaining part for the calculation of upper limits is done in the same way and the result 
is indicated in Fig.~\ref{UppLimOnBr90CL_NuhNumberBgSubtr_vs_mm2} by a thin smooth curve $(b)$.


To confirm the obtained results we end up with a simplified approach, less dependent on MC, where the background contribution 
is taken from a 5-th order polynomial approximation of the $m^{2}_{miss}$ distribution of the experimental data within the 
range of $m^{2}_{miss}>0.05$~GeV$^2/c^4$.
Again we do a common fit for signal plus background, where we allow the tuning of the background parameters for each investigated position.
In this case the obtained upper limit at 90\% CL  is shown with  dashed curve $(c)$ in Fig.~\ref{UppLimOnBr90CL_NuhNumberBgSubtr_vs_mm2}. 

Finally we select the result obtained with the MC background model described in subsection \ref{ss_SigSearchMCsubtr} as the main one, 
while the difference between three curves in Fig.~\ref{UppLimOnBr90CL_NuhNumberBgSubtr_vs_mm2} can be treated as a systematic
error of the method.


\subsection{Evaluation of the $|U_{\mu H}|^2$ mixing parameter upper limit}\label{SectComparison}\label{SigSearchMCbgSubtr}
\begin{figure}[!ht]
\begin{center}
\includegraphics[width=0.82\textwidth]{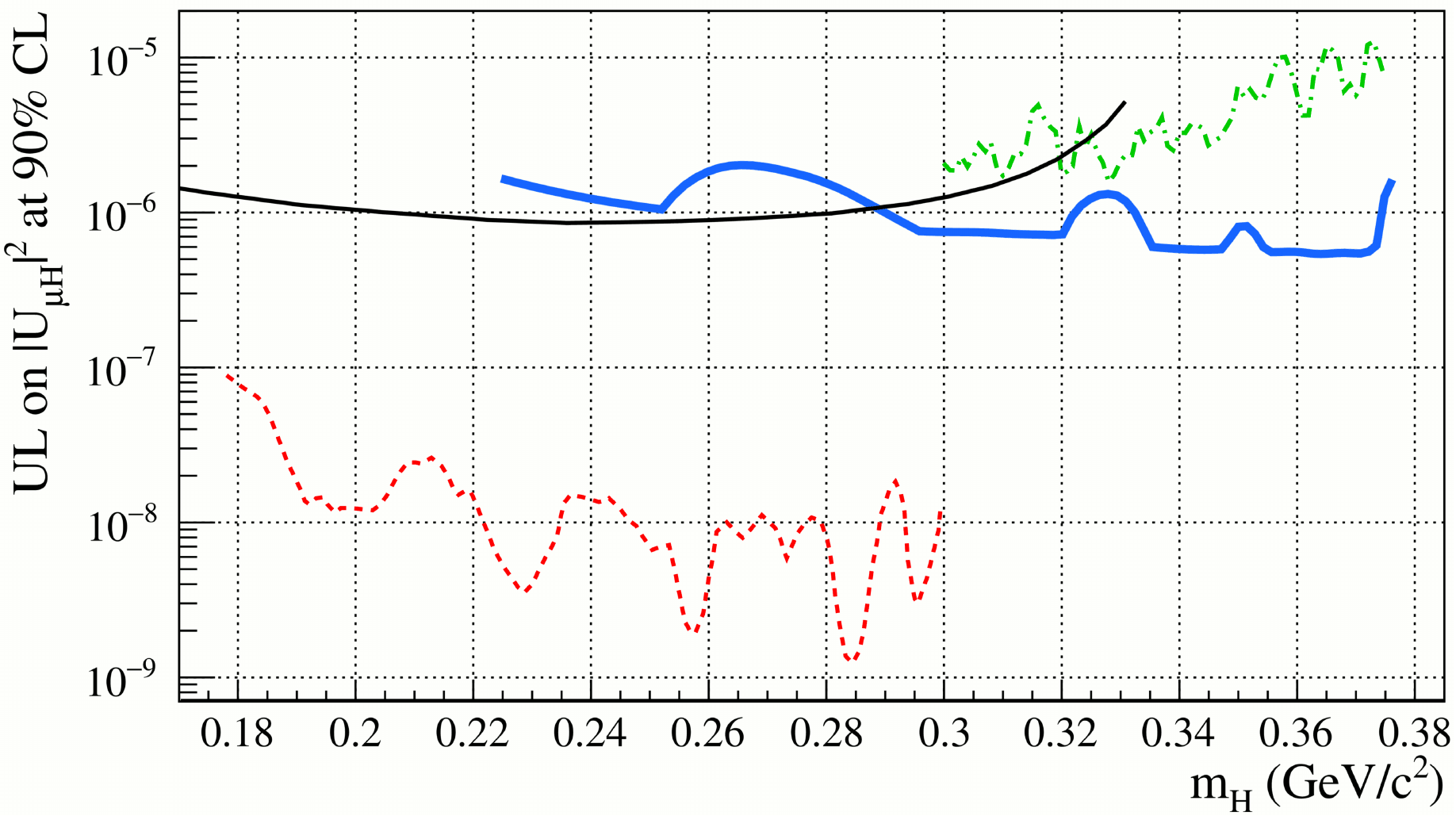} 
\caption[Upper limit on mixing matrix element $|U_{\mu H}|^{2}$ at 90\% CL in comparison with other experiments.]{
    \descrfnt
    The OKA upper limit on the mixing matrix element $|U_{\mu H}|^{2}$ at 90\% CL 
    is shown with solid blue curve, in comparison with preceding experiments:
    the black curve represents the result from KEK-E-089 \cite{KEK-E-089}, 
    the red dashed curve corresponds to the result from BNL-E949 \cite{ArtamonovEtAl-E949} obtained with stopped kaons,
    while the dash-dotted green curve indicates the recent result from CERN-NA62 \cite{PublNA62} 
}
\label{UL_U2_expComparison}
\end{center}
\end{figure}

Finally, we obtain an upper limit on the mixing parameter $|U_{\mu H}|^2$ between the muon neutrino 
and the heavy sterile neutrino $\nu_{H}$, see Fig.~\ref{UL_U2_expComparison}.
The coupling strength (see Fig.~\ref{FigHeavySterileNeutrinoKmuHnu}) between muon neutrino and $\nu_{H}$
is obtained from the relation:
$$
\frac{\Gamma(K \to \mu \nu_{H})}{\Gamma(K \to \mu \nu_{\mu})} = |U_{\mu H}|^{2} \cdot \lambda \cdot f_{\mathfrak{M}},
$$
where $\lambda$ is a kinematic factor, $f_{\mathfrak{M}}$ is a helicity factor in the matrix element, both 
arising for the case of massive neutrino in the final state \cite{ShrockCouplingUmH}. 

Since the obtained upper limit on $|U_{\mu H}|^{2}$ in the considered mass range 
does not exceed $10^{-5}$, the $\nu_{H}$ mean life time is estimated to be greater 
than $10^{-6}$~sec, assuming it decays to SM particles, \cite{PublNA62}. 
The corresponding $\nu_{H}$ mean flight distance, 
estimated with a MC simulation for the $\nu_{H}$ masses considered, ranges between 5--25~km.
Hence the heavy neutrino in our case, indeed, can be regarded as stable particle.

\section*{Conclusions}
The OKA 2012 data set was analyzed to search for 
heavy sterile neutrinos. A peak search method in the missing mass spectrum was used in the analysis. 
No signal is seen and the upper limit on the mixing between muon neutrino and a heavy sterile 
neutrino is set in the mass range 300--375~MeV/c$^{2}$.

Our result improves the old limit from KEK \cite{KEK-E-089} and the recent result from CERN NA62 \cite{PublNA62} 
for the mass range $m_{H}>290$ MeV/c$^{2}$, while at lower masses we still do not reach sensitivity of BNL-E949 \cite{ArtamonovEtAl-E949}.

\section*{Acknowledgements}
We express our gratitude to our colleagues in the accelerator department for the good performance of the U-70 during data taking; 
to colleagues from the beam department for the stable operation of the 21K beam line, including RF-deflectors, and to colleagues 
from the engineering physics department for the operation of the cryogenic system of the RF-deflectors.



\end{document}